\documentclass[twocolumn]{aastex6}

\begin{document}

\title{A universal correlation between star-formation activity and molecular gas properties across environments}


\author{Shuhei~Koyama\altaffilmark{1,2}, Yusei~Koyama\altaffilmark{3,4}, Takuji~Yamashita\altaffilmark{5}, Kana~Morokuma-Matsui\altaffilmark{1,6}, Hideo~Matsuhara\altaffilmark{1,7}, Takao~Nakagawa\altaffilmark{1}, Masao~Hayashi\altaffilmark{6}, Tadayuki~Kodama\altaffilmark{3,6,8}, Rhythm~Shimakawa\altaffilmark{9}, Tomoko~L.~Suzuki\altaffilmark{6}, Ken-ichi~Tadaki\altaffilmark{6}, Ichi~Tanaka\altaffilmark{3} and Moegi~Yamamoto\altaffilmark{4}}

\altaffiltext{1}{Institute of Space and Astronautical Science, Japan Aerospace Exploration Agency, 3-1-1 Yoshinodai, Chuo-ku, Sagamihara, Kanagawa 252-5210, Japan}
\altaffiltext{2}{Department of Physics, Tokyo Institute of Technology, 2-12-1 Ookayama, Meguro-ku, Tokyo 152-8551, Japan}
\altaffiltext{3}{Subaru Telescope, National Astronomical Observatory of Japan, National Institutes of Natural Sciences, 650 North A'ohoku Place, Hilo, HI 96720, U.S.A.}
\altaffiltext{4}{Department of Astronomical Science, Graduate University for Advanced Studies (SOKENDAI), 2-21-1 Osawa, Mitaka, Tokyo 181-8588, Japan}
\altaffiltext{5}{Research Center for Space and Cosmic Evolution, Ehime University, Bunkyo-cho, Matsuyama, Ehime 790-8577, Japan}
\altaffiltext{6}{National Astronomical Observatory of Japan, 2-21-1 Osawa, Mitaka, Tokyo 181-8588, Japan}
\altaffiltext{7}{Department of Space and Astronautical Science, Graduate University for Advanced Studies (SOKENDAI), 3-1-1 Yoshinodai, Chuo-ku, Sagamihara, Kanagawa 252-5210, Japan}
\altaffiltext{8}{Graduate School of Science, Tohoku University, 6-3 Aramaki Aza-Aoba, Sendai, Miyagi 980-8578, Japan}
\altaffiltext{9}{UCO/Lick Observatory, University of California, 1156 High Street, Santa Cruz, CA 95064, U.S.A.}

\begin{abstract}
We present the molecular gas mass fraction ($f_\mathrm{H_2}$) and star-formation efficiency (SFE) of local galaxies on the basis of our new CO($J=1-0$) observations with the Nobeyama 45m radio telescope, combined with the COLDGASS galaxy catalog, as a function of galaxy environment defined as the local number density of galaxies measured with SDSS DR7 spectroscopic data.
Our sample covers a wide range in the stellar mass and SFR, and covers wide environmental range over two orders of magnitude.
This allows us to conduct the first, systematic study of environmental dependence of molecular gas properties in galaxies from the lowest- to the highest-density environments in the local universe.
We confirm that both $f_\mathrm{H_2}$ and SFE have strong positive correlations with the SFR offset from the star-forming main sequence ($\Delta$MS), and most importantly, we find that these correlations are universal across all environments.
Our result demonstrates that star-formation activity within individual galaxies is primarily controlled by their molecular gas content, regardless of their global environment. 
Therefore, we claim that one always needs to be careful about the $\Delta$MS distribution of the sample when investigating the environmental effects on the H$_2$ gas content in galaxies.
\end{abstract}




\section{introduction} \label{sec:intro}

Galaxies reside in various environments such as clusters, groups or voids.
Galaxy properties are known to be strongly dependent on their surrounding environment.
In the local universe, the fraction of early-type galaxies increases with local galaxy density, while that of late-type galaxies decreases \citep[e.g.][]{Dre80, Got03}.
The average star-formation rate (SFR) of galaxies also decreases with increasing local galaxy density \citep[e.g.][]{Lew02, Gom03}.
These strong correlations between galaxy properties and environment suggest that the environment plays an important role in shaping the nature of individual galaxies.
It is believed that accelerated galaxy growth and/or efficient quenching of star formation in high-density environments could explain the environmental difference, but the physical causes (or mechanisms) responsible for the environmental effects are still unclear.

Many physical mechanisms have been proposed to explain the environmental trends, including interaction between galaxies and inter cluster/group medium (ICM), such as ram pressure stripping \citep{Gun72} or strangulation \citep{Lar80}, and galaxy--galaxy interaction such as galaxy harassment \citep{Moo98}, close encounter or major/minor merger \citep{Zab98}. 
Because all the proposed mechanisms are expected to influence the gas content in galaxies, it is very important to observe atomic (H\,{\sc i}) and molecular (H$_2$) gas content in galaxies to understand the environmental effects.
For H\,{\sc i} gas, many studies investigated the environmental impacts on the H\,{\sc i} gas content, showing that H\,{\sc i} gas fraction is decreased for star-forming galaxies in cluster environment \citep{Gio85, Chu09, Cor11, Ser12} and also in group environment \citep{Kil09, Ras12, Cat13, Bro17}.

These findings on the H\,{\sc i} deficiency for star-forming galaxies in high-density environments motivate us to study the environmental impacts on their H$_2$ gas content, which is the more direct fuel for star formation. 
If the H$_2$ gas is also reduced by the environmental effects (like H\,{\sc i} gas), then the average gas depletion time of those galaxies would become shorter, yielding more rapid consumption of H$_2$ gas in high-density environments.
In fact, there have been a number of studies investigating the environmental dependence of H$_2$ gas content with CO observations of nearby galaxy clusters.
Some pioneering work has shown that cluster galaxies have amounts of H$_2$ gas similar to those of the isolated galaxies \citep{Ken89, Cas91, Bos97, Lav98}.
These authors claimed that the environment does not affect the H$_2$ gas content in galaxies, because the H$_2$ gas is more strongly bound in the central part of galaxies by the gravitational potential than H\,{\sc i} gas.
However, some recent studies reported that cluster galaxies are deficient in H$_2$ gas content, suggesting that the molecular gas stripping takes place in cluster environment \citep[e.g.][]{Fum08, Cor12, Jab13, Sco13, Bos14}.
It is also reported that the cluster environment could enhance the H$_2$ gas mass fraction \citep[e.g.][]{Nak06, Mok16, Mok17}.
\citet{Mok16, Mok17} propose that the cluster environment could decrease the efficiency of star formation through molecular gas heating or turbulence at the same time, while some other studies suggest an {\it increased} star formation efficiency due to the gas compression in cluster galaxies \citep[e.g.][]{Ebe14, Lee17}.

In these ways, our understanding of the environmental impacts on the H$_2$ gas content (and star-formation efficiency) in galaxies is still highly uncertain.
The different results (or discrepancy) between different studies probably originate in the different sample selection, different environmental coverage, or insufficient sample size.
In particular, most of the previous studies mentioned above attempted to compare the results on cluster galaxies and field galaxies. 
However, galaxy morphologies and the average star-formation rates are known to change monotonically with environment. A more systematic study covering a wide environmental range from low- to high-density environment is needed. 

It is also important to carefully select the targets for molecular gas (CO) observations. 
Typical star-forming galaxies exhibit a tight positive correlation between stellar mass ($M_*$) and SFR; so-called ``main sequence (MS)" of star-forming galaxies \citep{Dad07, Elb07, Noe07}, and it has been shown that both H$_2$ gas mass fraction ($f_\mathrm{H_2} = M_\mathrm{H_2}/M_*$) and star-formation efficiency ($SFE = SFR/M_\mathrm{H_2}$) are strongly dependent on their location in the SFR--M$_*$ diagram. Based on the extensive CO galaxy observations of nearby galaxies (COLDGASS survey), \citet{Sai12} demonstrated that both $f_\mathrm{H_2}$ and SFE significantly increase with the SFR ``offset'' values from the star-forming main sequence ($\Delta$MS), implying that star-formation activity is controlled by the molecular gas content in galaxies. 
It is therefore very important to pay extra attention to the $\Delta$MS range of the sample; otherwise the results can easily be biased due to the different $\Delta$MS range between different samples.

In this paper, we present our new CO observations of nearby star-forming galaxies with the Nobeyama 45m radio telescope (NRO~45m) to cover wide range in $\Delta$MS and environment.
By combining our NRO~45m data with the literature data, 
this work provides the first systematic study on the environmental impacts on the H$_2$ gas content in galaxies across a wide environmental range.

This paper is organized as follows. In Section~\ref{sec:data}, we describe our sample selection and our new CO observations with NRO~45m telescope, as well as our supplementary data from the COLDGASS survey. In Section~\ref{sec:results}, we present our main results on the environmental independence of the H$_2$ gas mass fraction and SF efficiency at fixed $\Delta$MS, and we compare our results with previous studies. Finally, our conclusions are presented in Section~\ref{sec:conclusion}.
Throughout this paper, we assume the $\Lambda$CDM universe with $H_0 = 70~\mathrm{km~s^{-1}~Mpc^{-1}}$, $\Omega_m = 0.3$ and $\Omega_{\Lambda} = 0.7$, and \citet{Kro01} initial mass function (IMF).

\section{data} \label{sec:data}

\subsection{Definition of Environment}
In this work, we define the galaxy environment based on the local number density of galaxies calculated by \citet{Mat17}, who computed the local density of 738,143 galaxies selected from SDSS DR7 spectroscopic data \citep{Aba09} within the area of $105^\circ < R.A. < 270^\circ$ and $-5 < DEC. < 75^\circ$.
The local number density of each galaxy ($\Sigma_5$) is calculated using the projected distance to the fifth-nearest neighbor galaxy within a velocity window of $\pm 1000$ km s$^{-1}$, which corresponds to a redshift slice of $\Delta z = \pm0.003$, and is expressed as:
\begin{equation}
\Sigma_5 = \frac{5}{\pi D_{p,5}^2}\,\mathrm{[Mpc^{-2}]},
\end{equation}
where $D_{p,5}$ is the projected comoving distance to the fifth-nearest neighbor galaxy.
To take into account the redshift dependence on the completeness limit of SDSS spectroscopic survey, $\Sigma_5$ is normalized by the median density measured within the same velocity window:
\begin{equation}
\rho_5 = \frac{\Sigma_5}{\langle\Sigma_5\rangle}.
\end{equation}

\begin{figure*}[tbp]
	\begin{center}
		\includegraphics[width=170mm]{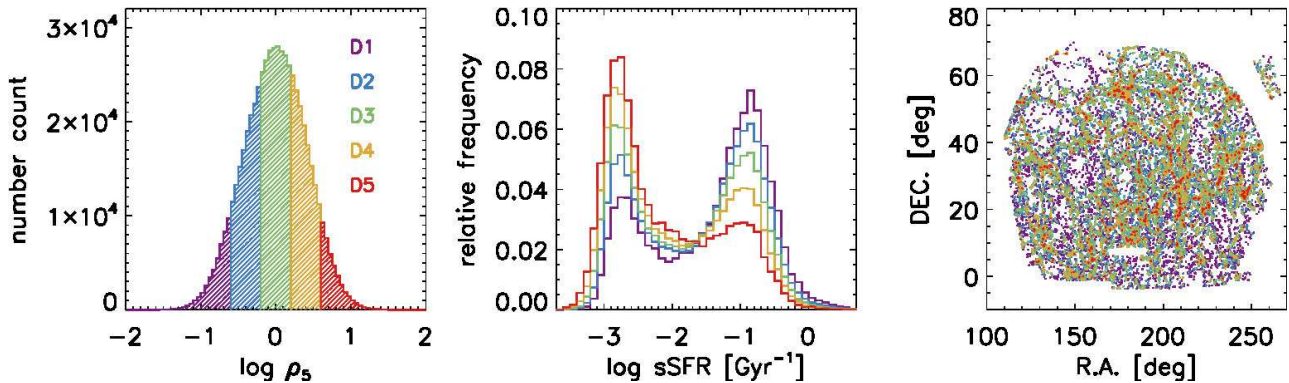}
	\end{center}
	\caption{(left): The distribution of $\rho_5$ for all the SDSS DR7 sample within $105^\circ < R.A. < 270^\circ$ and $-5 < DEC. < 75^\circ$ computed by \citet{Mat17}. In this paper, we define the environment based on $\rho_5$, and our definition of D1--D5 environment bins are shown with different color shades in this diagram.
	(middle): The relative frequency of sSFR for each environmental bin. 
	(right): The spatial distribution of the sample galaxies at $z=0.06-0.07$ on the sky.
	In all the panels, the color coding indicates different environmental bin (redder colors indicate higher-density environment).}
	\label{density}
\end{figure*}

Figure~\ref{density} (left) shows the density distribution for all the SDSS sample in the above R.A./Dec. range.
Our sample covers roughly two orders of magnitude in terms of the local galaxy density.
As shown in Figure~\ref{density} (left), we divide the sample into five environmental bins (D1--D5) based on the following criteria: $\log\rho_5 \le -0.6$ for D1, $-0.6 < \log\rho_5 \le -0.2$ for D2, $-0.2 < \log\rho_5 \le 0.2$ for D3, $0.2 < \log\rho_5 \le 0.6$ for D4, $0.6 < \log\rho_5$ for D5.
The majority of galaxies are located in D3 bin, while the galaxies in D1 and D5 bins are rare.
Figure~\ref{density} (middle) shows the distribution of specific SFR ($sSFR = SFR/M_*$) of galaxies for each environmental bin.
This plot demonstrates that our definition of environment works reasonably well, and the relative fraction of passive and star-forming galaxies clearly changes with environment. 
Figure~\ref{density} (right) shows the spatial distribution of galaxies at $z=0.06-0.07$ on the sky, to visually demonstrate that our local density measurement can robustly trace from rich clusters as well as the filamentary large-scale structures to low-density fields.
We note that the previous studies are equivalent to the comparison between D3 and D5 bins.

\subsection{Sample Selection}
\label{sample selection}

The goal of this paper is to understand the environmental impact on the H$_2$ gas content in galaxies across a wide environmental range.
Ideally, the sample should be selected uniformly from D1--D5. 
Using the SDSS galaxies for which the local density measurements are available, we select the target galaxies for the CO observation with NRO~45m telescope based on the following criteria:

1) To select SF galaxies, we restrict the sample to those satisfying $EW_\mathrm{H\alpha} > 4 \mathrm{\AA}$, and adopt the SF galaxy criteria defied by \citet{Kew01} on the BPT diagram \citep{Bal81}. Here, we also require signal-to-noise ratio (S/N) $>$3 for the four major lines (H$\alpha$, H$\beta$, [NII]6584, [OIII]5007) to compute the line flux ratios.

2) We select the galaxies with $z \ge 0.03$ so that the major part of galaxies are covered with a single beam size of NRO~45m telescope.
The beam size of NRO~45m at 110~GHz is 15~arcsec, which corresponds to 10~kpc at $z=0.03$.

3) We select the galaxies with $L_{IR} \ge 10^{11}\ \mathrm{L_\odot}$ (i.e.\ with Luminous Infrared Galaxies (LIRGs) class luminosity), in order to complement the COLDGASS sample which mostly covers the galaxies on and below the SF main sequence (see Section 2.4 for details).
The $L_{IR}$s are computed by \citet{Koy15} following \citet{Tak10} using \textit{AKARI} far-IR photometric bands \citep{Kaw07}; WIDE-S (90~$\mu$m) and WIDE-L (140~$\mu$m).

We find 1,568 galaxies satisfying the above criteria, from which we finally select 32 targets (6--7 galaxies for each D1--D5 bin) considering the visibility from the NRO~45m telescope.
The properties of our NRO~45m target galaxies are summarized in Table~\ref{SP}.
We note that our sample does not necessarily cover very rich cluster environments, but we stress again that our main aim here is to ``bridge'' previous studies focusing on cluster and field environments, by applying a continuous environmental coverage from low- to high-density environments quantified with the local galaxy density.

\subsection{Observation and Data Reduction} \label{subsec:obs}

We performed $^{12}$CO($J=1-0$) observations of 32 galaxies during the period from December to February in the 2015/2016 semester using the NRO~45m telescope (CG151005: Y. Koyama et al.).
The CO emission line of the rest-frame 115.271\,GHz is shifted to 99.372 -- 111.914\,GHz according to the redshifts of our sample.
We used a two-beam, two-polarization, sideband-separating SIS receiver, TZ \citep{Nak13}, and a copy of a part of the FX-type correlator for the Atacama Compact Array, SAM45 \citep{Kam12}.
Typical on-source integration time was 1.5 hours for each galaxy.
During the observation, the pointing accuracy was checked every hour by observing SiO maser sources at 43\,GHz.
The image rejection ratio (IRR) was measured for each target and for each observing date, and the IRR values were 5 -- 30\,dB at the center of the intermediate frequency.
The system noise temperature ($T_{\rm sys}$) of our observation ranges between 100 -- 540\,K with the median $T_{\rm sys}$ value of 150\,K.

The flux calibration was performed by the chopper wheel method and the measured IRR scaling factor ($f$):
\begin{equation}
f = 1 + \frac{1}{10^{IRR/10}}.
\end{equation}
The main beam temperature ($T_{\rm mb}$) was calculated by $T_{\rm mb} = T^{*}_a / \eta_{\rm mb}$, where $T^{*}_a$ is the antenna temperature.
The main beam efficiency ($\eta_{\rm mb}$) during the semester was 0.44 according to the NRO website\footnote{\url{http://www.nro.nao.ac.jp/~nro45mrt/html/prop/eff/eff_latest.html}}.

\begin{figure*}[p]
	\begin{center}
		\includegraphics[width=170mm]{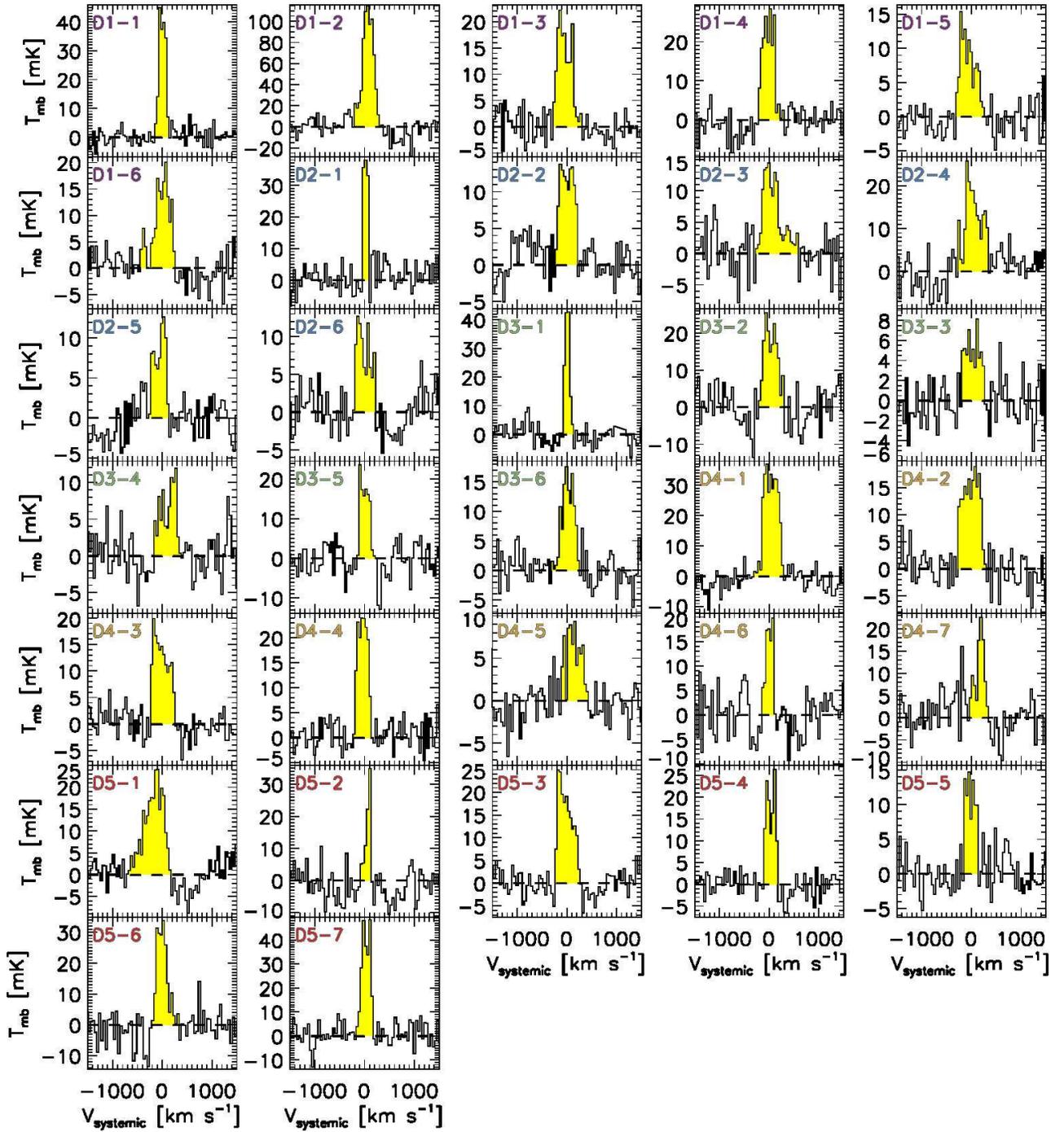}
	\end{center}
	\caption{The CO spectra of galaxies derived with our NRO~45m observations. The CO flux of each galaxy is calculated by summing the intensity within the velocity range colored in yellow. Their physical properties are summarized in Table~\ref{SP}.}
	\label{spectra}
\end{figure*}

Data reduction was performed by using the NEWSTAR software, which was developed by NRO based on the Astronomical Image Processing System (AIPS) package.
We used the observed data with wind velocities of $<5\,\rm m\,s^{-1}$, pointing accuracy better than $5^{\prime\prime}$,  $T_{\rm sys}$ better than 300\,K, and IRR values larger than 7\,dB.
We also extracted the data with the rms noise temperature levels of ($T_{\rm rms}$) $> 0.045\,\rm K$ in the $T_{\rm mb}$ scale at a velocity resolution of 200\,km\,s$^{-1}$ to exclude bad baseline spectra.
Then, we subtracted baselines by linear fitting and sum up for both polarizations.
After binning up to 40\,km\,s$^{-1}$ resolution, we calculated the integrated intensity $I_{\rm CO}$ according to $I_{\rm CO} = \int T_{\rm mb}\,dv$.
The error in $I_{\rm CO}$ was calculated according to $T_{\rm rms}\sqrt{\Delta V_{\rm e}\Delta v}$, where $\Delta V_e$ is the full line width of CO spectra.
The $T_{\rm rms}$ at the velocity resolution of $\Delta v = 40\,\rm km\,s^{-1}$ is 2.1--8.5\,mK($T_{\rm mb}$) for our targets.
We detected CO emission line from all the target galaxies with the signal (peak temperature) -to-noise (rms) ratio (S/N) of 3 -- 17. Finally, we calculate the CO luminosity ($L^{\prime}_{\rm CO}$) based on the following equation:
\begin{equation}
L'_\mathrm{CO} = \frac{\Omega_b I_\mathrm{CO} D^2_{L}}{(1+z)^3}~\mathrm{[K~km~s^{-1}~pc^2]},
\end{equation}
where $\Omega_{\rm b}$ is the beam solid angle of $\theta_{\rm mb}$, and $D_{L}$ is the luminosity distance.
The observational results are summarized in Table 2, and the observed CO spectra are shown in Figure~\ref{spectra}.

\begin{table*}[p]
	\caption{Summary of physical quantities of our NRO~45m sample. The asterisks in the SFR column denote the galaxies for which we replaced their SFRs with the SFRs derived from $L_{IR}$ we computed from AKARI FIR photometry (see text). We assign 0.3-dex uncertainties for these two sources, which corresponds to the typical error of the {\it AKARI} WIDE-L band photometry.}
	\label{SP}
\hspace{-1cm}
\resizebox{\textwidth}{!}{
		\begin{tabular}{cccccccccc} \hline
SDSS ID & Env. & $\log\rho_5$ & $z$ & $\log SFR\,\mathrm{[M_\odot\,yr^{-1}]}$ & $\log M_*\,\mathrm{[M_\odot]}$ & $\log L_{IR}\,\mathrm{[L_\odot]}$ & $I_{CO}\,\mathrm{[K\,km\,s^{-1}]}$ & $\log L_{CO}\,\mathrm{[K\,km\,s^{-1}pc^2]}$ & $\log M_\mathrm{H_2}\,\mathrm{[M_\odot]}$ \\ \hline \hline
J082656.14+040503.5 & D1-1 & -1.14 & 0.056 &  0.75$^{+ 0.24}_{- 0.19}$ & 10.77$^{+ 0.10}_{- 0.08}$ & 11.23 &  7.29$\pm$ 0.27 &  9.36$\pm$ 0.09 & 10.00$\pm$ 0.09\\
J133223.99+110620.4 & D1-2 & -1.13 & 0.031 &  0.67$^{+ 0.18}_{- 0.09}$ & 10.03$^{+ 0.15}_{- 0.10}$ & 11.19 & 28.83$\pm$ 1.27 &  9.48$\pm$ 0.10 & 10.11$\pm$ 0.10\\
J103009.77+093505.8 & D1-3 & -0.99 & 0.084 &  1.30$^{+ 0.10}_{- 0.09}$ & 10.89$^{+ 0.09}_{- 0.09}$ & 11.50 &  6.26$\pm$ 0.40 &  9.63$\pm$ 0.15 & 10.26$\pm$ 0.15\\
J082919.82+061744.8 & D1-4 & -0.96 & 0.048 &  1.27$^{+ 0.04}_{- 0.05}$ & 10.68$^{+ 0.10}_{- 0.10}$ & 11.06 &  7.06$\pm$ 0.45 &  9.23$\pm$ 0.15 &  9.86$\pm$ 0.15\\
J083948.87+263402.7 & D1-5 & -0.84 & 0.069 &  0.88$^{+ 0.14}_{- 0.14}$ & 10.50$^{+ 0.08}_{- 0.13}$ & 11.20 &  4.26$\pm$ 0.35 &  9.31$\pm$ 0.19 &  9.94$\pm$ 0.19\\
J140757.00+321249.2 & D1-6 & -0.83 & 0.087 &  0.94$^{+ 0.17}_{- 0.14}$ & 10.61$^{+ 0.09}_{- 0.09}$ & 11.62 &  5.49$\pm$ 0.40 &  9.60$\pm$ 0.17 & 10.24$\pm$ 0.17\\
J153907.86+140123.0 & D2-1 & -0.57 & 0.038 &  0.89$^{+ 0.13}_{- 0.10}$ & 10.40$^{+ 0.11}_{- 0.09}$ & 11.02 &  4.26$\pm$ 0.31 &  8.82$\pm$ 0.17 &  9.45$\pm$ 0.17\\
J102449.68+053327.1 & D2-2 & -0.46 & 0.067 &  0.77$^{+ 0.39}_{- 0.19}$ & 10.68$^{+ 0.13}_{- 0.09}$ & 11.30 &  4.64$\pm$ 0.31 &  9.32$\pm$ 0.15 &  9.95$\pm$ 0.15\\
J140132.89+305242.3 & D2-3 & -0.43 & 0.072 &  1.07$^{+ 0.11}_{- 0.13}$ & 10.69$^{+ 0.10}_{- 0.09}$ & 11.44 &  5.18$\pm$ 0.59 &  9.42$\pm$ 0.26 & 10.05$\pm$ 0.26\\
J122104.98+113752.3 & D2-4 & -0.40 & 0.068 &  1.44$^{+ 0.21}_{- 0.14}$ & 10.87$^{+ 0.16}_{- 0.10}$ & 11.81 &  7.85$\pm$ 0.69 &  9.56$\pm$ 0.20 & 10.19$\pm$ 0.20\\
J095804.35+124414.3 & D2-5 & -0.27 & 0.062 &  1.10$^{+ 0.04}_{- 0.11}$ & 10.07$^{+ 0.11}_{- 0.10}$ & 11.26 &  2.93$\pm$ 0.35 &  9.06$\pm$ 0.27 &  9.69$\pm$ 0.27\\
J095545.86+314840.1 & D2-6 & -0.21 & 0.084 &  1.34$^{+ 0.09}_{- 0.10}$ & 11.03$^{+ 0.11}_{- 0.10}$ & 11.49 &  3.37$\pm$ 0.39 &  9.36$\pm$ 0.27 & 10.00$\pm$ 0.27\\
J091528.96+202754.5 & D3-1 & -0.18 & 0.068 &  1.02$^{+ 0.19}_{- 0.16}$ & 10.72$^{+ 0.10}_{- 0.08}$ & 11.22 &  5.03$\pm$ 0.32 &  9.36$\pm$ 0.15 & 10.00$\pm$ 0.15\\
J155725.27+244332.1 & D3-2 & -0.08 & 0.041 &  0.79$^{+ 0.18}_{- 0.12}$ & 10.57$^{+ 0.09}_{- 0.08}$ & 11.05 &  6.76$\pm$ 0.67 &  9.08$\pm$ 0.23 &  9.71$\pm$ 0.23\\
J102420.51+035911.9 & D3-3 & -0.02 & 0.099 &  1.30$^{+ 0.22}_{- 0.13}$ & 11.00$^{+ 0.13}_{- 0.11}$ & 11.61 &  2.19$\pm$ 0.28 &  9.31$\pm$ 0.30 &  9.95$\pm$ 0.30\\
J101410.56+342034.7 & D3-4 &  0.02 & 0.038 &  1.00$^{+ 0.39}_{- 0.12}$ & 10.34$^{+ 0.14}_{- 0.12}$ & 11.20 &  3.17$\pm$ 0.43 &  8.67$\pm$ 0.31 &  9.30$\pm$ 0.31\\
J115801.99+103136.2 & D3-5 &  0.09 & 0.065 &  1.15$^{+ 0.12}_{- 0.13}$ & 10.61$^{+ 0.10}_{- 0.09}$ & 11.56 &  4.31$\pm$ 0.52 &  9.26$\pm$ 0.28 &  9.89$\pm$ 0.28\\
J095835.38+004434.0 & D3-6 &  0.12 & 0.065 &  1.35$^{+ 0.14}_{- 0.12}$ & 10.80$^{+ 0.09}_{- 0.09}$ & 11.34 &  4.63$\pm$ 0.37 &  9.29$\pm$ 0.19 &  9.92$\pm$ 0.18\\
J110319.65+151202.4 & D4-1 &  0.23 & 0.136 &  1.56$^{+ 0.30}_{- 0.30}$* & 11.04$^{+ 0.10}_{- 0.10}$ & 11.32 & 12.35$\pm$ 0.54 & 10.32$\pm$ 0.10 & 10.95$\pm$ 0.10\\
J110326.80+161820.8 & D4-2 &  0.36 & 0.068 &  1.11$^{+ 0.12}_{- 0.09}$ & 11.04$^{+ 0.13}_{- 0.09}$ & 11.20 &  7.22$\pm$ 0.47 &  9.53$\pm$ 0.15 & 10.16$\pm$ 0.15\\
J123932.01+272950.4 & D4-3 &  0.37 & 0.057 &  1.22$^{+ 0.12}_{- 0.10}$ & 10.86$^{+ 0.09}_{- 0.09}$ & 11.21 &  6.15$\pm$ 0.32 &  9.31$\pm$ 0.12 &  9.94$\pm$ 0.12\\
J103422.30+442349.2 & D4-4 &  0.37 & 0.052 &  0.83$^{+ 0.09}_{- 0.10}$ & 10.43$^{+ 0.09}_{- 0.10}$ & 11.31 &  6.20$\pm$ 0.32 &  9.24$\pm$ 0.12 &  9.87$\pm$ 0.12\\
J101632.22+085842.9 & D4-5 &  0.40 & 0.103 &  1.10$^{+ 0.13}_{- 0.10}$ & 10.61$^{+ 0.09}_{- 0.08}$ & 11.70 &  2.81$\pm$ 0.37 &  9.45$\pm$ 0.31 & 10.08$\pm$ 0.31\\
J103014.76+110415.8 & D4-6 &  0.41 & 0.065 &  1.08$^{+ 0.14}_{- 0.10}$ & 10.81$^{+ 0.10}_{- 0.09}$ & 11.32 &  3.28$\pm$ 0.42 &  9.14$\pm$ 0.29 &  9.77$\pm$ 0.29\\
J142343.69+064441.3 & D4-7 &  0.59 & 0.157 &  1.30$^{+ 0.17}_{- 0.16}$ & 10.89$^{+ 0.10}_{- 0.10}$ & 11.05 &  3.72$\pm$ 0.47 &  9.91$\pm$ 0.29 & 10.54$\pm$ 0.29\\
J090105.72+343526.2 & D5-1 &  0.61 & 0.065 &  1.96$^{+ 0.30}_{- 0.30}$* & 10.21$^{+ 0.09}_{- 0.11}$ & 11.84 &  9.55$\pm$ 0.57 &  9.61$\pm$ 0.14 & 10.25$\pm$ 0.14\\
J102601.79+213537.6 & D5-2 &  0.64 & 0.042 &  0.66$^{+ 0.12}_{- 0.10}$ & 10.41$^{+ 0.09}_{- 0.08}$ & 11.01 &  3.14$\pm$ 0.41 &  8.76$\pm$ 0.30 &  9.40$\pm$ 0.30\\
J095725.12+004351.3 & D5-3 &  0.67 & 0.087 &  1.48$^{+ 0.13}_{- 0.12}$ & 11.02$^{+ 0.10}_{- 0.08}$ & 11.46 &  7.27$\pm$ 0.43 &  9.73$\pm$ 0.14 & 10.36$\pm$ 0.14\\
J144352.34+162827.2 & D5-4 &  0.67 & 0.054 &  1.04$^{+ 0.16}_{- 0.11}$ & 10.61$^{+ 0.11}_{- 0.08}$ & 11.25 &  4.80$\pm$ 0.25 &  9.15$\pm$ 0.12 &  9.78$\pm$ 0.12\\
J090509.86+172557.9 & D5-5 &  0.76 & 0.067 &  1.11$^{+ 0.11}_{- 0.10}$ & 10.68$^{+ 0.09}_{- 0.09}$ & 11.32 &  3.23$\pm$ 0.29 &  9.16$\pm$ 0.21 &  9.80$\pm$ 0.20\\
J151604.07+195134.5 & D5-6 &  0.83 & 0.049 &  1.34$^{+ 0.17}_{- 0.09}$ & 10.74$^{+ 0.11}_{- 0.09}$ & 11.21 &  7.82$\pm$ 0.71 &  9.29$\pm$ 0.21 &  9.92$\pm$ 0.21\\
J142306.89+204324.0 & D5-7 &  1.14 & 0.049 &  0.83$^{+ 0.19}_{- 0.17}$ & 10.91$^{+ 0.09}_{- 0.09}$ & 11.22 &  9.05$\pm$ 0.49 &  9.34$\pm$ 0.13 &  9.98$\pm$ 0.13\\
\hline
		\end{tabular}
		}
\end{table*}

\subsection{COLDGASS Sample} \label{subsec:CG}

In this study, we also use the CO data of nearby galaxies publicly available from the COLDGASS survey \citep{Sai11, Sai12}. 
The COLDGASS is one of the most extensive extragalactic CO surveys performed with the IRAM~30m telescope for 366 galaxies in the redshift range of $0.025 < z < 0.05$.
 
Because the COLDGASS sample is drawn from SDSS, we can measure their local density in the same way as our sample.
We note that the COLDGASS survey is not originally designed to cover a wide environmental range, and so it tends to lack galaxies in the high- and low-density environments (such as D1 or D5 in our definition).
In addition, most of their samples are located on or below the MS relation, and the number of galaxies above the MS is limited.
Our NRO~45m sample is therefore complementary to the COLDGASS sample with these respects. 

We only use the COLDGASS sample with CO detection with $S/N > 5$, for 169 of which we have the local galaxy density measurement.
We find that one target in our NRO~45m sample overlaps with one of the COLDGASS sample, and we confirmed that our CO intensity measurement is consistent with the COLDGASS measurement (with the IRAM~30m) within the error (typically 20\,\%) for this source.

\subsection{The Final Sample}

By combining our new NRO~45m sample and the COLDGASS sample, our final sample includes 203 galaxies in total.
The redshift range is $z=0.025-0.16$.
We note that there is a possibility that $\rho_5$ does not probe the physically same environment across this redshift range.
Because, $\rho_5$ may vary depending on the stellar mass limit used for density calculation, particularly when the stellar mass distribution depends on the environment.
In this study, we used all galaxies selected from a relatively wide redshift range to have a sufficiently large sample size across all environments,
however we confirmed that our conclusions do not change even if we restrict the sample to those within a narrow redshift range (e.g. $z=0.025-0.05$).  

We calculate the molecular gas mass ($M_\mathrm{H_2}$) with $M_\mathrm{H_2} = \alpha_\mathrm{CO}L'_\mathrm{CO}$, where $\alpha_\mathrm{CO}$ is the CO-to-H$_2$ conversion factor.
We adopt the Galactic value of $\alpha_\mathrm{CO} = 4.3$ M$_\odot\ (\mathrm{K\ km\ s^{-1}\ pc^2})^{-1}$,
which includes the contribution of heavy elements (mainly from helium), as commonly used in studies of star-forming galaxies in the local universe \citep{Bol13}.
Here we assume the fixed $\alpha_\mathrm{CO}$ value as many studies, although the $\alpha_\mathrm{CO}$ may be different in starburst (i.e.\ high $L_{\rm IR}$) galaxies \citep[e.g.][]{Sol97}.
However, we verify that our results on the environmental dependence are not affected even if we adopt the variable conversion factors.

For the final sample, we use SFR and M$_*$ from the SDSS value-added catalog by Max Plank Institute for Astrophysics and Johns Hopkins University (MPA/JHU) group \footnote{\url{http://wwwmpa.mpa-garching.mpg.de/SDSS/DR7/}} for all galaxies.
We use SFR$_\mathrm{SDSS}$ for all the galaxies, although we originally selected the NRO~45m sample based on L$_\mathrm{IR}$ (see Section~\ref{sample selection}),
because it is not possible to estimate L$_\mathrm{IR}$ for many of the COLDGASS sample with low SFR due to the detection limit of the \textit{AKARI} FIR data.
There may be a small systematic offset between SFR$_\mathrm{IR}$ and SFR$_\mathrm{SDSS}$ (by $\sim$0.3-dex level), as reported by some recent studies \citep[e.g.][]{Lee13}, but the results presented in this study based on {\it relative} comparison between different environments is not affected.

As exception, we replace SFR$_\mathrm{SDSS}$s with the SFR$_\mathrm{IR}$s (see Table~\ref{SP}) regarding two of our (FIR-detected) NRO~45m targets.
They show unrealistically small SFR$_\mathrm{SDSS}$s; their SFR$_\mathrm{SDSS}$s are two orders of magnitude smaller than SFR$_\mathrm{IR}$ because of the wrong SDSS fiber position.
We note that our results do not change even if we exclude these two galaxies from our analysis.

\begin{figure*}[tbp]
	\begin{center}
		\includegraphics[width=170mm]{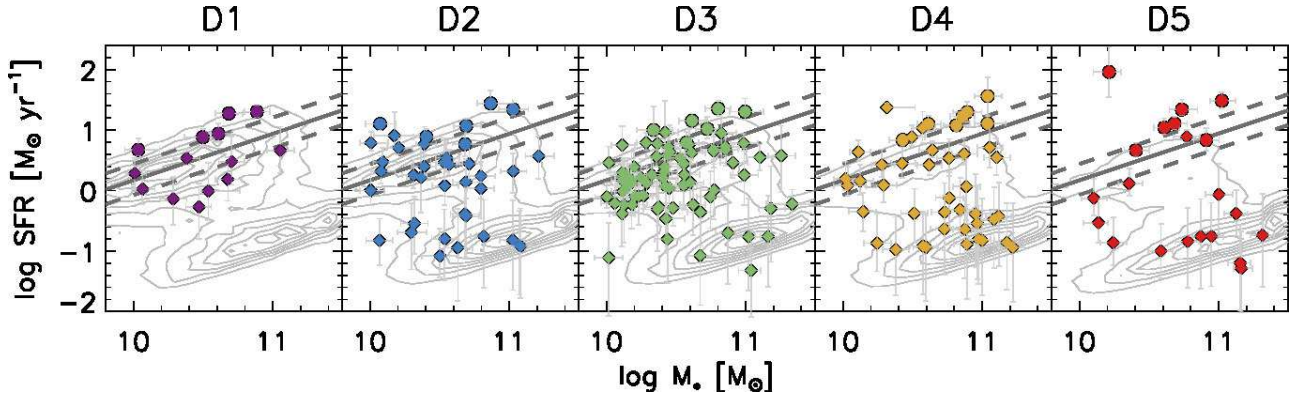}
	\end{center}
	\caption{SFR--M$_*$ diagram for each environment (D1--D5).
		The filled circles show our NRO45m sample, and the filled diamonds show the sample from COLDGASS with CO detection.
		The solid line drawn in all the panels shows the main sequence relation in the local universe reported by \citet{Elb07}, and the broken lines show its 1-$\sigma$ scatter.
		The gray contours show the distribution of all the SDSS sample for each environmental bin.}
	\label{M_SFR}
\end{figure*}

Figure~\ref{M_SFR} shows the distribution of the final sample on the M$_*$--SFR plane in each environmental bin.
As shown in this plot, the COLDGASS samples (diamonds) are distributed mainly below the MS relation, while our NRO galaxies (circles) are distributed on/above the MS relation.
The distribution of all the SDSS samples significantly changes with environment (see gray contours in each panel), but we stress again that the aim of this work is to cover a wide range in M$_*$, SFR, and environment, and to test the environmental dependence of galaxy properties at a fixed position on the M$_*$--SFR plane. 

\section{Results \& Discussion} \label{sec:results}

\subsection{$\mathrm{M_{H_2}}$--SFR Relation in Different Environments} \label{subsec:KS low}

\begin{figure*}[tbp]
	\begin{center}
		\includegraphics[width=170mm]{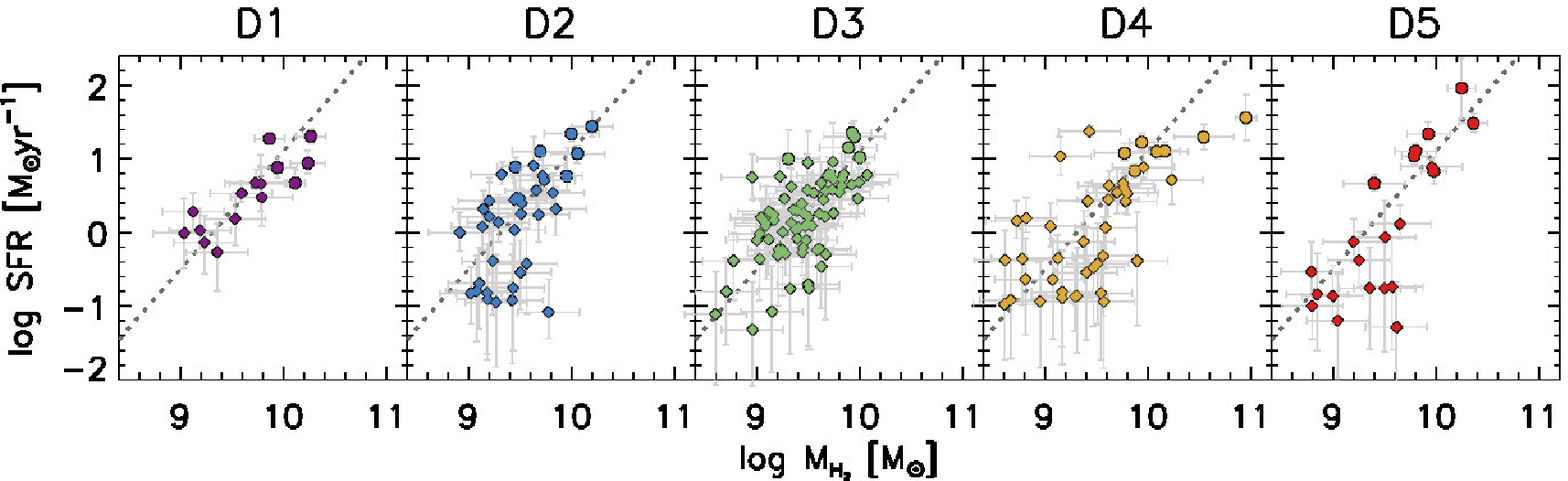}
	\end{center}
	\caption{The relation between $\mathrm{M_{H_2}}$ and SFR for each environmental bin.
		For comparison, we show the best-fitting result for D3 bin as gray dotted lines in all the panels. The meanings of the symbols are the same as Figure~\ref{M_SFR}.
		}
	\label{SFR_MH2}
\end{figure*}

First, we investigate the environmental dependence of the correlation between $\mathrm{M_{H_2}}$ and SFR, which is the fundamental correlation describing star formation \citep{Ken98}. 
Figure~\ref{SFR_MH2} shows the $\mathrm{M_{H_2}}$--SFR relation for galaxies in each environmental bin.
To compare the distribution in different environments, we first fit the data points in the D3 bin (i.e.\ the most ``general'' environment) by the liner regression form of $y=A+Bx$, where {\it y} is substituted by $\log SFR/\mathrm{M_\odot~yr^{-1}}$ and {\it x} is substituted by $\log M_\mathrm{H_2}/\mathrm{M_\odot}$, by using the Ordinary Least-Squares (OLS)--Bisector method.
The gray dotted line in each panel of Figure~\ref{SFR_MH2} shows the best-fitting result for D3 bin.
We then measure the distance in the orthogonal direction to this best-fitted line for each galaxy, and perform the two-sample Kolmogolov-Smirnov (KS) test between D1/D2/D4/D5 and D3 samples KS plots are shown in Appendix \ref{Ap}).
The derived p-values, the probabilities that the D1/D2/D4/D5 and D3 samples are drawn from the same parent population, are summarized in Table~\ref{KStest}.
Their p-values are $>$0.1 in all cases, suggesting that we cannot rule out the hypothesis that $\mathrm{M_{H_2}}$--SFR relations are different between different environments.
Therefore, we conclude that the $\mathrm{M_{H_2}}$--SFR relation does not significantly depend on the environment; i.e.\ the star-formation law is universal across environments.

\subsection{Environmental Independence of $f_\mathrm{H_2}$ and SFE at fixed $\Delta$MS} \label{subsec:deltaMS}

As mentioned in Section~\ref{sec:intro}, it is reported that $f_\mathrm{H_2}$ and SFE are strongly correlated with $\Delta$MS.
In this subsection, we study the $\Delta$MS--$f_\mathrm{H_2}$ and $\Delta$MS--SFE correlation for each environmental bin, to test the environmental dependence of the $f_\mathrm{H_2}$ and SFE at fixed $\Delta$MS. 
Our aim here is to understand whether or not galaxies having the same star-formation activity, but existing in different environments, have the same amount of H$_2$ gas and are forming stars with the same efficiency.
It is well established that the MS relation changes with redshift, but there is no measurable change within the redshift range of our sample ($z=0.025-0.16$) \citep{Spe14}.
Furthermore, the MS relation does not significantly change with the environment as reported by many recent studies \citep{Pen10, Koy13}, and thus we simply define the $\Delta$MS for all galaxies in our sample as the SFR offset value from the MS at $z=0$:

\begin{equation}
\Delta\mathrm{MS} = \log SFR - \log SFR_{MS},
\end{equation}
where $SFR_{MS}$ denotes the SFR of galaxies on the MS relation at a given M$_*$.
We adopt the MS relation in the local universe defined by \citet{Elb07} as follows:
\begin{equation}
\log SFR_{MS} = 0.77\log M_* - 7.53,
\end{equation}
as shown with the solid line in each panel of Figure~\ref{M_SFR}.
We note that our conclusions are unchanged even if we separately define the MS for each environment. 

\begin{figure*}[tbp]
	\begin{center}
		\includegraphics[width=170mm]{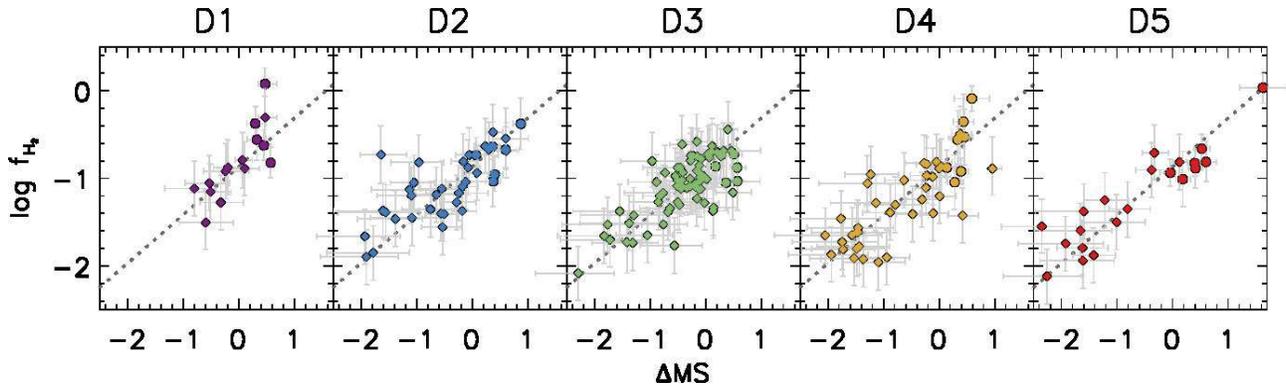}
	\end{center}
	\caption{$\Delta$MS--$f_\mathrm{H_2}$ relation for each environmental bin.
		The meanings of the color codings and symbol styles are the same as in Figure~\ref{SFR_MH2}.}
	\label{deltaMS_FH2}
\end{figure*}

\begin{figure*}[tbp]
	\begin{center}
		\includegraphics[width=170mm]{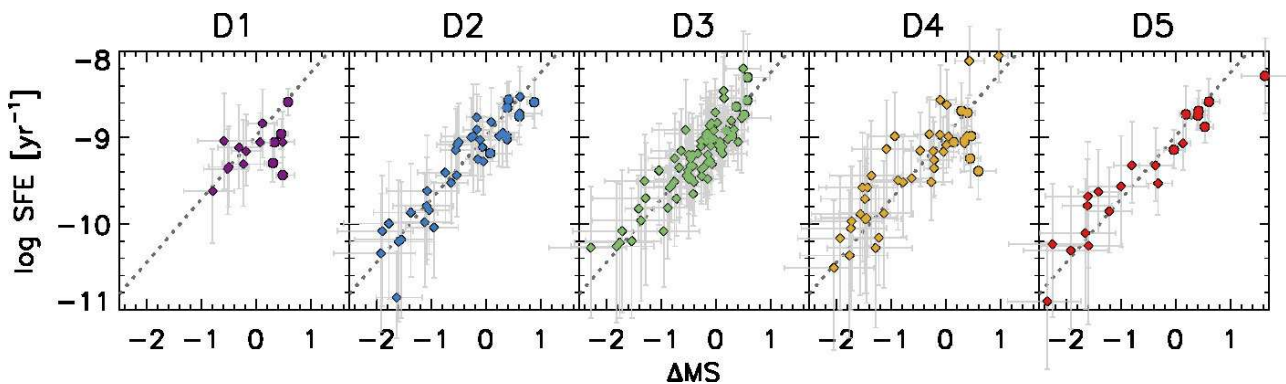}
	\end{center}
	\caption{The same plot as Fig.~\ref{deltaMS_FH2} but for SFE.  
		The meanings of the color coding and symbols are the same as in Figure~\ref{SFR_MH2} and Figure~\ref{deltaMS_FH2}.}
	\label{deltaMS_SFE}
\end{figure*}

We show the $\Delta$MS--$f_\mathrm{H_2}$ and the $\Delta$MS--SFE correlation for each environmental bin (D1--D5) in Figure~\ref{deltaMS_FH2} and Figure~\ref{deltaMS_SFE}, respectively. 
It can be seen that there exist strong $\Delta$MS--$f_\mathrm{H_2}$ and $\Delta$MS--SFE correlations in all environmental bins, and more importantly, there is no significant difference between different environments.
As we did in Figure~\ref{SFR_MH2}, we first fit the data points in D3 bin by a linear regression form of $y=A+Bx$, where {\it y} is substituted by $\log f_\mathrm{H_2}$ (and $\log SFE$), and {\it x} is substituted by the $\Delta$MS.
We then perform the KS tests for the residual distributions as we did in the previous section (KS plots are shown in Appendix \ref{Ap}), and we confirm that their p-values are $>$0.1 in all cases as summarized in Table~\ref{KStest},
suggesting that we cannot rule out the hypothesis that all the environmental subsamples are drawn from the same parent population. 

Overall, our results revealed that galaxies having the same star-formation activity ($\Delta$MS), but existing in different environments, have the same amount of H$_2$ gas and are forming stars with the same efficiency.
This is the most important result of this study, and this is the answer to the original question raised at the beginning of this section. 

The environmental independence of the $\Delta$MS--$f_\mathrm{H_2}$ (or --SFE) correlations suggest that 
most galaxies are located on the same $\Delta$MS--$f_\mathrm{H_2}$ (or --SFE) correlation, with very few galaxies being scattered from the sequence.
This implies that galaxies (in all environments) must evolve by moving along this sequence; i.e.\ if a galaxy loses a fraction of its molecular gas content (due to some environmental effects), the star-formation activity needs to be ``adjusted'' accordingly in relatively short time scale to come back onto the original $\Delta$MS--$f_\mathrm{H_2}$ (or --SFE) correlation. 

\begin{table}[htbp]
	\caption{The p-values derived from the KS test for the residual distribution, where the residuals are defined as the orthogonal distances from the best-fitted line for D3 bin to each galaxy on the $\mathrm{M_{H_2}}$--SFR, $\Delta$MS--$f_\mathrm{H_2}$ and $\Delta$MS--SFE planes. } 
	\label{KStest}
	\begin{center}
		\begin{tabular}{cccc} \hline
& $\mathrm{M_{H_2}}$--SFR & $\Delta$MS--$f_\mathrm{H_2}$ & $\Delta$MS--SFE \\
\hline \hline
D1 & 0.58 &  0.50 & 0.11 \\
D2 & 0.39 & 0.91 & 0.96 \\
D3 & 1.00 & 1.00 & 1.00 \\
D4 & 0.15 & 0.79 & 0.35 \\
D5 & 0.12 & 0.48 & 0.94 \\
\hline
		\end{tabular}
	\end{center}
\end{table}

\subsection{Comparison with Previous Studies}

Many studies have discussed the environmental impacts on the molecular gas content in galaxies, mainly by comparing CO data of cluster and field galaxies as described in Section~\ref{sec:intro}. We recall that the goal of this study is to extend and complement these earlier studies by performing a CO galaxy survey by covering a wide environmental range. In this study, we achieved this goal by combining our new CO data from NRO~45m observations (Section~\ref{subsec:obs}) and COLDGASS survey (Section~\ref{subsec:CG}), and we demonstrated that the $\Delta$MS--$f_\mathrm{H_2}$ and $\Delta$MS--$f_\mathrm{H_2}$ correlations known for field galaxies \citep[e.g.][]{Sai12} are universal across all environments. Here, we provide some implications for the environmental impacts on the H$_2$ gas content in galaxies based on our findings, as well as on previous studies.

As we mentioned in Section~\ref{sec:intro}, many studies have investigated the environmental effects on the molecular gas content in various types of galaxies using CO observations. 
For example, \citet{Ken89}, \citet{Cas91} and \citet{Lav98} performed CO observations of nearby cluster galaxies selected with B-band or far-IR luminosities.
These studies suggest that the H$_2$ gas content ($f_\mathrm{H_2}$) in galaxies is the same between cluster and isolated galaxies, but it should be noted that the B-band and far-IR selections are sensitive to the current/recent star formation, so that their samples tend to be biased to actively star-forming galaxies. 
On the other hand, more recent studies by \citet{Fum08}, \citet{Sco13} and \citet{Bos14} showed that cluster galaxies tend to have {\it lower} H$_2$ gas mass fraction on average than isolated galaxies. Their galaxy samples are optically selected, which tends to be less biased to recent star formation.  

In our current work, we demonstrated that the $f_\mathrm{H_2}$ is constant at fixed $\Delta$MS in all environments, suggesting that the average $f_\mathrm{H_2}$ of each environmental subsample should be determined by the average $\Delta$MS of the sample.
Because the original $\Delta$MS distribution for {\it all} galaxies is shown to be significantly dependent on environment (see the middle panel of Figure~\ref{density} or the contour of Figure~\ref{M_SFR}), it is not surprising to see the trend that $f_\mathrm{H_2}$ (or SFE) tends to be lower in higher-density environments, if the galaxies are selected randomly from each environment without considering their $\Delta$MS distribution. We therefore speculate that the apparently different suggestions made by different authors regarding the environmental impacts on the molecular gas content in galaxies can be interpreted as being due to their different sample selection. 
We emphasize again that the H$_2$ gas content in galaxies is the main driver of star formation in galaxies regardless of their global environment, and so our results may leave an important message for all studies discussing the environmental effects on the molecular gas content in galaxies; i.e.\ one always needs to be careful about how one selects the sample when investigating the environmental effects on the H$_2$ gas content in galaxies.

We finally comment that, although we attempted to cover as wide an environmental range as possible, our sample does not fully cover the highest-density rich cluster environment. Our ``D5'' environment seems to correspond to a relatively wide environmental range from clusters to the filamentary structures (as demonstrated in Figure~\ref{density}-right). Recent studies by \citet{Mok16, Mok17} demonstrate that the spiral galaxies in the Virgo cluster have {\it larger} amount of $\mathrm{M_{H_2}}$ and {\it longer} gas depletion time ($M_\mathrm{H_2}/SFR$) than field/group galaxies, claiming that the cluster environment affects the efficiency of star formation through molecular gas heating or turbulence. On the other hand, some studies suggest that ram pressure in rich cluster environments compresses the gas in galaxies, and can enhance the star formation in the cluster member galaxies \citep[e.g.][]{Ebe14, Lee17}. In these ways, the effect of rich cluster environments on molecular gas content in galaxies is still under debate, but it is possible that galaxies in extremely rich environments could deviate from the $\Delta$MS--$f_\mathrm{H_2}$ (or $\Delta$MS--SFE) relations. A more comprehensive study, which hopefully covers a wider environmental range with larger samples from the lowest- to highest-density environments, is needed to fully understand the effect of such extreme environments.

\section{Conclusion} \label{sec:conclusion}

In this paper, we present the environmental effects on the H$_2$ gas mass fraction ($f_\mathrm{H_2}$) and star-formation efficiency (SFE) of local galaxies, based on our new CO(1-0) observation at the NRO~45m telescope and the data from COLDGASS survey.
Our final CO(1-0) sample covers a wide stellar mass and SFR range, and also covers a wide environmental range over two orders of magnitudes in the local number density of galaxies measured with SDSS DR7 spectroscopic galaxy sample.
Our findings are summarized as follows.

1) The correlation between $\mathrm{M_{H_2}}$ and SFR, the so-called ``integrated Kennicutt-Schmidt law'', does not significantly depend on their surrounding environment; i.e.\ the star-formation law is universal across environments.

2) Both $f_\mathrm{H_2}$ and SFE show strong positive correlations with $\Delta$MS, consistent with recent studies, and these correlations are universal in all environments. This result suggests that the star-formation process occurring within individual galaxies is not strongly affected by their global environment, but primarily controlled by their H$_2$ gas content. The tight $\Delta$MS--$f_\mathrm{H_2}$ (and $\Delta$MS--SFE) correlation implies that galaxies (in all environments) must evolve by moving along the same sequence. 

3) The environmentally independent $\Delta$MS--$f_\mathrm{H_2}$ correlation revealed in this study leaves an important message for all studies discussing environmental impacts on the molecular gas properties in galaxies. We emphasize that the sample selection, more specifically the $\Delta$MS range of the samples, is critically important for any environmental studies focusing on the H$_2$ gas in galaxies. To perform a fair comparison between the samples drawn from different environments, the sample should be carefully selected so that there is no bias in the $\Delta$MS range between the samples.

\acknowledgments
This research is based on SDSS-III, observations at the Nobeyama Radio Observatory (NRO) and observations with {\it AKARI}, a JAXA project with the participation of ESA.

We thank the members of NRO for their observational support, and thank Yasuhiro Matsuki for his offer of the SDSS environment calculated sample.

Funding for SDSS-III has been provided by the Alfred P. Sloan Foundation, the Participating Institutions, the National Science Foundation, and the U.S. Department of Energy Office of Science. The SDSS-III web site is \url{http://www.sdss3.org/}.

SDSS-III is managed by the Astrophysical Research Consortium for the Participating Institutions of the SDSS-III Collaboration including the University of Arizona, the Brazilian Participation Group, Brookhaven National Laboratory, University of Cambridge, Carnegie Mellon University, University of Florida, the French Participation Group, the German Participation Group, Harvard University, the Instituto de Astrofisica de Canarias, the Michigan State/Notre Dame/JINA Participation Group, Johns Hopkins University, Lawrence Berkeley National Laboratory, Max Planck Institute for Astrophysics, Max Planck Institute for Extraterrestrial Physics, New Mexico State University, New York University, Ohio State University, Pennsylvania State University, University of Portsmouth, Princeton University, the Spanish Participation Group, University of Tokyo, University of Utah, Vanderbilt University, University of Virginia, University of Washington, and Yale University.

This work was financially supported in part by a Grant-in-Aid for the Scientific Research (No.\,26800107, 26247030) by the Japanese Ministry of Education, Culture, Sports and Science.



\vspace{5mm}
\facilities{NRO\,45m}
\software{NEWSTAR}



\appendix
\section{The cumulative functions for the KS test} \label{Ap}
\begin{figure}[tbp]
	\begin{center}
		\includegraphics[width=170mm]{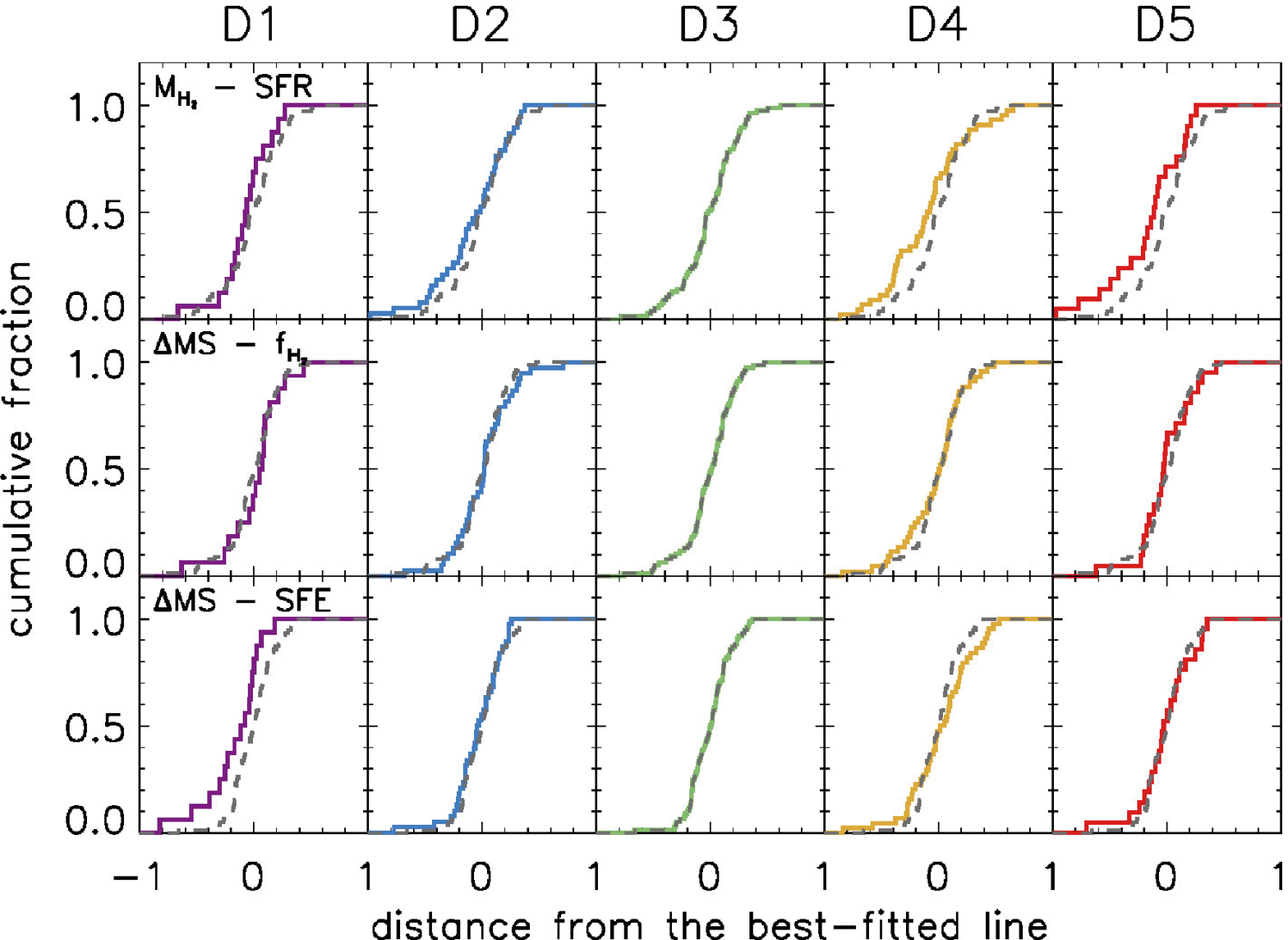}
	\end{center}
	\caption{
		The cumulative functions for each environment (D1--D5).
		From top to bottom panel, we show the cumulative functions corresponding to Figure \ref{SFR_MH2}, \ref{deltaMS_FH2} and \ref{deltaMS_SFE}.
		The solid lines show the cumulative functions for each environmental bin, and the broken lines show that for D3 bin for comparison. Because we measure the distance from the best-fitted line defined for D3 sample, the solid line and broken lines are identical for D3 panels by definition.}
	\label{cumulative}
\end{figure}

In Section~\ref{sec:results}, we use the KS test to investigate the environmental dependence of the relation between molecular gas related properties and star-formation activity.
In the KS tests, we measure the distance in the orthogonal direction to the best-fitted line defined in D3 bin for each environmental bin, and compare the cumulative functions among D1/D2/D4/D5 and D3 samples. In Figure \ref{cumulative}, we show the cumulative functions to visually demonstrate our conclusion that there is no significant difference among different environments.

\end{document}